# DosimeTron: Automating Personalized Monte Carlo Radiation Dosimetry in PET/CT with Agentic AI

**Research paper**

**Eleftherios Tzanis, PhD[1,2*], Michail E. Klontzas, MD, PhD[1,2,5], Antonios Tzortzakakis MD, PhD[1,3,4]**

1. Division of Radiology, Department for Clinical Science, Intervention and Technology (CLINTEC), Karolinska Institutet, Stockholm, Sweden

2. Artificial Intelligence and Translational Imaging (ATI) Lab, Department of Radiology, School of Medicine, University of Crete, Heraklion, Crete, Greece

3. Department of Nuclear Medicine and Medical Physics, Section Nuclear Medicine Huddinge, Karolinska University Hospital, Stockholm, Sweden

4. Department of Nuclear Medicine and Medical Physics, Theranostics Trial Center, Karolinska University Hospital, Stockholm, Sweden

5. Computational Biomedicine Laboratory, Institute of Computer Science Foundation for Research and Technology Hellas (ICS - FORTH), Heraklion, Crete, Greece

*Corresponding author

**Eleftherios Tzanis, PhD**

E-mail: eleftherios.tzanis@ki.se

ORCID: 0000-0003-0353-481X

## Abstract

**Background:** Internal radiation dosimetry in PET/CT is predominantly performed using population-averaged S-value methods that cannot account for inter-patient anatomical variability. Patient-specific Monte Carlo (MC) dosimetry is the recognised gold standard but has not achieved clinical adoption due to workflow complexity and prohibitive computation times.

**Purpose:** To develop and evaluate DosimeTron, an agentic AI system for automated patient-specific MC internal radiation dosimetry in PET/CT examinations.

**Materials and Methods:** In this retrospective study, DosimeTron was evaluated on a publicly available PSMA-PET/CT dataset comprising 597 studies from 378 male patients acquired on three scanner models ($^{18}$F, n = 369; $^{68}$Ga, n = 228). The system uses GPT-5.2 as its reasoning engine and 23 tools exposed via four Model Context Protocol servers, automating DICOM metadata extraction, image preprocessing, MC simulation, organ segmentation, and dosimetric reporting through natural-language interaction. Agentic performance was assessed using diverse prompt templates spanning single-turn instructions of varying specificity and multi-turn conversational exchanges, monitored via OpenTelemetry traces. Dosimetric accuracy was validated against OpenDose3D across 114 cases and 22 organs using Pearson's r, Lin's concordance correlation coefficient (CCC), and Bland-Altman analysis.

**Results:** Across all prompt templates and all runs, no execution failures, pipeline errors, or hallucinated outputs were observed. Pearson's r ranged from 0.965 to 1.000 (median 0.997; all $p < 0.001$) and CCC from 0.963 to 1.000 (median 0.996). Mean absolute percentage difference was below 5% for 19 of 22 organs (median 2.5%). Total per-study processing time was 32.3 ± 6.0 minutes.

**Conclusion:** DosimeTron autonomously executed complex dosimetry pipelines across diverse prompt configurations and achieved high dosimetric agreement with OpenDose3D at clinically acceptable processing times, demonstrating the feasibility of agentic AI for patient-specific Monte Carlo dosimetry in PET/CT.

## 1. Introduction

PET/CT represents a fundamental component of modern nuclear medicine, providing simultaneous functional and anatomical information that has transformed diagnostic accuracy across clinical domains such as oncology, cardiology and neurology [1]. More recently, the same molecular targeting principles underlying diagnostic PET have been extended to radiopharmaceutical therapy, establishing the theranostics paradigm [2]. Across clinical PET/CT applications, accurate personalized radiation dosimetry is essential for patient safety and procedure optimization.

Radiation dosimetry in PET/CT has traditionally relied on pre-calculated S-values, as formalized through the Medical Internal Radiation Dose (MIRD) schema [3]. S-values represent the absorbed dose delivered to a target organ per unit cumulated activity in a source organ and are derived from mathematical or voxelized anthropomorphic phantoms representing the average person. This methodology is implemented in widely used software tools such as OLINDA/EXM [4]. While straightforward, this approach is inherently population-based and cannot account for inter-patient variability in organ geometry, anatomical positioning, or radiotracer distribution, which may differ substantially between individuals and introduce significant dosimetric uncertainty [5, 6].

Patient-specific Monte Carlo (MC) simulation is widely regarded as the reference standard for personalized internal radiation dosimetry: a voxelized phantom is constructed directly from the patient's CT, the PET image serves as the three-dimensional activity map, and radiation transport is simulated through the individual's specific anatomy and radiopharmaceutical distribution, yielding a patient-specific three-dimensional dose distribution [7]. Despite its dosimetric superiority, MC dosimetry has not yet been widely adopted in routine clinical practice. The workflow requires extensive data preprocessing and expertise in radiation transport simulation, while the extensive computation times can still limit incorporation in busy clinical workflows.

Recent advances in large language model (LLM)-based agentic systems have demonstrated their capacity to autonomously orchestrate complex workflows across a range of medical imaging and clinical decision-support tasks [8-11]. The aim of this study was to develop an agentic system, DosimeTron, designed to automate end-to-end personalised MC internal radiation dosimetry for patients undergoing PET/CT examinations. The agent communicates with the clinical expert in natural language: the user provides the patient's scan and specifies the requested analysis, ranging from a complete dosimetry pipeline to targeted processing of specific PET/CT data, and the agent autonomously executes the appropriate sequence of operations. These include DICOM metadata extraction, image preprocessing, voxelised phantom construction, MC simulation, organ segmentation, and dosimetric reporting across anatomical structures. Hardware resources are managed dynamically by the agent to parallelize computation, reducing processing times without requiring programming expertise or manual user configuration. In essence, the system addresses a translational gap between advances in MC-based dosimetry and their limited adoption in clinical practice by automating the complex, multi-step and time-intensive processes involved. By eliminating manual preprocessing, manual MC simulation setup, and iterative analysis, the framework enables clinically feasible patient-specific radiation dosimetry, facilitating its integration into routine nuclear medicine workflows.

## 2. Materials and Methods

Data were obtained retrospectively from a publicly available open-access dataset [12]. No institutional ethics committee approval was required. The study was conducted in accordance with the CLAIM reporting guidelines [13].

### 2.1 Agentic System Architecture

DosimeTron was developed for end-to-end patient-specific MC radiation dosimetry in PET/CT, employing GPT-5.2 as its core reasoning engine. The system's computational toolkit was built using the Model Context Protocol (MCP) [14], an open standard defining a language-agnostic client-server interface between LLM-powered agents and external tool implementations. Four Python-based MCP servers were developed, responsible for DICOM metadata extraction, image preprocessing, MC simulation, and organ dose extraction, collectively exposing 23 tools to the agent. A desktop application (Electron v28.3.3) serves as the MCP host, launching and managing the four servers and aggregating their tools. Communication between host and servers uses standard input/output streams with JSON-RPC 2.0 framing. Agent behavior, pipeline execution logic, and state-tracking rules are encoded in a structured system prompt. The agent operates through a reasoning loop in which it observes the current pipeline state, issues tool calls, and updates the state upon receiving the results (Figure 1). Each MCP server is stateful, maintaining in-memory objects such as image volumes, processed arrays, dose maps, and metadata, across tool calls, enabling multi-step workflows without serialization through the LLM context.

### 2.2 Graphical User Interface

For efficient communication between the agentic system and the user, a desktop application was developed using the Electron framework (v28.3.3). The interface provides a natural-language chat panel through which the user initiates the pipeline and provides parameters, an MCP server management panel for selective activation of individual MCP servers and their tools, and a dashboard displaying active servers, available tools, and real-time token usage.

## 2.3 Monte Carlo Radiation Dosimetry

Prior to simulation, the agent extracts all required metadata from the PET DICOM headers, including radionuclide identity, injected activity, injection and acquisition timestamps, patient weight, and pixel calibration parameters. The CT volume is resampled onto the PET grid using Lanczos interpolation [15]. PET pixel values (Bq/mL or SUV_bw) are converted to absolute activity in MBq per voxel. Multi-planar CT-PET fusion images are generated for alignment verification. Both volumes are reoriented to RAI (Right-Anterior-Inferior) convention and saved as Gate-compatible MetaImage files (CTRS.mhd, ACTM.mhd).

MC simulation is performed using GATE v9.4, built on the Geant4 toolkit [16]. The agentic system automatically generates the complete set of input files: a voxelized patient phantom is constructed by mapping Hounsfield unit values to tissue-specific densities and elemental compositions, the activity map defines the spatially distributed radioactive source with voxel emission probability proportional to measured PET activity, the full decay cascade of the identified radionuclide is simulated using the 'emstandard_opt4' physics list [17] with production thresholds of 1 mm in air and 0.1 mm in the patient volume. Absorbed dose is recorded voxel-by-voxel via a DoseActor with mass-weighting, yielding dose, energy deposition, mass, and statistical uncertainty maps. The simulation is distributed across multiple CPU cores as specified by the user, with one independent Gate process per core launched in parallel Docker containers, reducing computation time proportionally to the number of cores employed.

## 2.4 Organ Dose Extraction and Reporting

The raw Gate dose is scaled to the true absorbed dose as $D_{\text{true}}(r) = D_{\text{gate}}(r) \times \tilde{A}/N_{\text{sim}}$, where $\tilde{A}$ is the cumulated activity (Bq·s) and $N_{\text{sim}}$ is the total number of simulated primary decays. Organ/tissue segmentation is then performed on the original CT using TotalSegmentator [18]. The resulting binary masks are spatially co-registered to the scaled dose distribution and applied as region-of-interest filters. The agentic system provides the following dosimetric quantities for each organ/tissue:

**Dose Conversion Factor (DCF)** - derived directly from the MC output as the mean absorbed dose per simulated primary decay, independently of any kinetic model:

$$DCF = \frac{D_{gate,organ}}{N_{sim}} \; [Gy/Bq \cdot s]$$

The DCF encodes the spatial transport of radiation through the patient-specific anatomy. Users with organ-specific cumulated activities $\tilde{A}_{\text{custom}}$ from time-activity curve analyses or compartmental pharmacokinetic models can compute absorbed dose as $D_{\text{organ}} = \tilde{A}_{\text{custom}} \times \text{DCF}$ without repeating the simulation, rendering the methodology kinetic-model agnostic.

**Dose rate at scan acquisition time** - computed as:

$$\dot{D}_{organ}(t_{scan}) = DCF \times A_{organ}(t_{scan}) \; [Gy/s]$$

where $A_{\text{organ}}(t_{\text{scan}})$ is the organ activity at scan acquisition time, obtained by summing the calibrated PET voxel values over the organ mask.

**Organ absorbed dose** - obtained by time-integrating the activity to determine $\tilde{A}$. With $\lambda_{\text{phys}} = ln\, 2/T_{\text{phys}}$, two integration modes are provided:

$$\tilde{A}_{scan} = \frac{A_{(t_{scan})}}{\lambda_{phys}}, \qquad \tilde{A}_{inj} = \frac{A_{(t_{scan})} e^{\lambda_{phys} \cdot \Delta t}}{\lambda_{phys}}$$

representing conservative (scan→∞) and upper-bound (injection→∞) estimates respectively, where $\Delta t = t_{\text{scan}} - t_{\text{inj}}$. When organ-specific biological half-lives $T_{\text{bio}}$ are provided by the user, the effective decay constant $\lambda_{\text{eff}} = \lambda_{\text{phys}} + ln\, 2/T_{\text{bio}}$ replaces $\lambda_{\text{phys}}$, yielding $\tilde{A}_{\text{eff}} = A(t_{\text{scan}})/\lambda_{\text{eff}}$ with effective half-life $T_{\text{eff}} = T_{\text{phys}} \cdot T_{\text{bio}}/(T_{\text{phys}} + T_{\text{bio}})$. For each anatomical structure, the system reports mean dose, maximum dose, volume, mass, statistical uncertainty, and dose-volume histogram (DVH) data.

## 2.5 Dataset

The pipeline was evaluated on an open-access PSMA-PET/CT dataset from the Cancer Imaging Archive [12], comprising 597 whole-body PET/CT studies from 378 male patients (mean age 71.5 ± 8.1 years, range 48–90 years), acquired between 2014 and 2022 on three scanner models: Siemens Biograph 64-4R TruePoint (116 studies), Siemens Biograph mCT Flow 20 (251 studies), and GE Discovery 690 (230 studies). Two radionuclides, $^{18}F$ (369 studies) and $^{68}Ga$ (228 studies), are represented within the dataset.

## 2.6 Experimental Evaluation Protocol

Seven test prompts were designed across two categories. Five single-turn prompts (A1-A5) required the agent to execute the complete pipeline autonomously using a single instruction, varying in specificity from exhaustive step enumeration (A1) to a minimal plain-language instruction (A4), and covering scan-to-infinity and injection-to-infinity integration modes, $T_{eff}$ correction, dose rate analysis, and a range of primary counts. Two multi-turn prompts (B1-B2) involved request-response exchanges, between the user and the agent, to assess pipeline state retention, step-wise execution, and stateful recall across turns. B2 included a robustness test in which incorrect metadata was provided to assess whether the agent deferred to DICOM-extracted values. Selected cases were repeated under different configurations to assess the consistency of system responses.

## 2.7 Pipeline Monitoring and Observability

To monitor system behaviour, a tracing system was implemented using the OpenTelemetry framework [19] across the host and all four MCP servers. Each execution is recorded as a hierarchical trace: a top-level 'AGENT' span contains individual 'TOOL' spans capturing tool name, server routing, input arguments, output, status, and duration. LLM messages and tool call sequences are recorded following the OpenInference semantic convention [20]. Traces are exported live to Arize Phoenix [21]. This infrastructure enabled verification of correct server and tool selection, argument passing, and result propagation for each test run.

## 2.8 Radiation Dosimetry Validation

To evaluate dose accuracy, 114 cases were randomly selected from the dataset and processed independently using OpenDose3D [22], a validated manual radiation dosimetry framework supporting local energy deposition, FFT convolution, and Monte Carlo simulation. The Monte Carlo mode was employed for direct methodological correspondence. Both pipelines used the same DICOM input and TotalSegmentator for anatomical structure segmentation. Organ/tissue dose rates produced by OpenDose3D were compared against those generated by the agentic pipeline. Agreement between the two methods was assessed per organ using Pearson's correlation coefficient (r), Lin's concordance correlation coefficient (CCC) [23], Bland-Altman analysis (bias and 95% limits of agreement) [24], and the mean absolute percentage difference (MAPD). All analyses were performed separately for $^{18}F$ and $^{68}Ga$ subgroups and for the combined cohort.

All experiments were conducted on a workstation equipped with an Intel Core i9-14900 CPU, 128 GB RAM, and an NVIDIA RTX 5090 GPU (32 GB VRAM). All dosimetric results and the time efficiency analysis are based on runs configured with $10^7$ primaries distributed over 28 parallel CPU cores.

# 3. Results

## 3.1 Dosimetric Output

In response to the different user prompts, DosimeTron autonomously extracted all required dosimetric metadata from PET DICOM headers, preprocessed CT and PET imaging data, performed patient-specific Monte Carlo radiation dosimetry, executed automatic segmentation of organs and tissues, and generated verification figures (Figure 2) enabling the user to confirm correct CT-PET registration and activity map calibration. For each study, a CSV report was generated containing per-organ/tissue dosimetric quantities, including dose rates at scan acquisition time, absorbed doses under scan-to-infinity and injection-to-infinity time integration, and dose conversion factors for use with custom kinetic models. Three-dimensional dose distributions were computed and visualised as dose overlays on the corresponding CT anatomy for inspection of the spatial dose distribution within the patient body (Figure 3). Despite the imaging data being acquired across three different scanner models and containing heterogeneous DICOM metadata, the system successfully identified the radionuclide and extracted all required dosimetric parameters from all studies. Tables 1 and 2 present the mean dosimetric quantities for the 22 anatomical structures which received the highest radiation doses for the $^{18}$F and $^{68}$Ga cases respectively.

## 3.2 Validation Against OpenDose3D

Scan-time dose rates produced by OpenDose3D [22] were compared against those generated by the agentic pipeline across 114 cases and 22 anatomical structures (Table 3). Pearson's r ranged from 0.965 to 1.000 (median 0.997, $p < 0.001$) and Lin's concordance correlation coefficient (CCC) from 0.963 to 1.000 (median 0.996). Bland-Altman analysis showed negligible systematic bias (range -0.022 to +0.030 μGy/s) and narrow limits of agreement (LoA) for the majority of organs. The MAPD was below 5% for 19 of 22 organs (median 2.5%). Higher percentage differences were observed for the adrenal glands (5.2-5.9%) and thyroid (5.5%). However, the corresponding absolute differences in dose rate were ≤ 0.006 μGy/s for these structures, translating to clinically negligible differences in absorbed dose. Agreement was consistent across radionuclides, with median r of 0.997 and median CCC of 0.995 for $^{18}F$ and 0.998 and 0.997 for $^{68}Ga$. Bland-Altman and scatter plots for the eight organs receiving the highest absorbed doses are shown in Figures 4 and 5, respectively.

### 3.3 Computational Performance

The total per-study processing time of the proposed agentic pipeline was 32.3 ± 6.0 min (median 28.7 min, range 23.8-44.0 min). DICOM metadata extraction and image preprocessing together required approximately 12 seconds, while post-simulation processing (organ segmentation, dose extraction, and report generation) took 4.9 ± 0.6 min. The MC simulation constituted the dominant computational step, accounting for approximately 84% of total execution time (27.0 ± 6.7 min). In comparison, the mean per-study MC simulation time for OpenDose3D was 411.7 ± 89.7 min (median 358.3 min, range 315.4–588.6 min).

### 3.4 Agentic Behaviour and Robustness

All agentic runs were monitored through OpenTelemetry traces. Each of the seven prompt templates was applied repeatedly across different studies and with varying simulation parameters, spanning single-turn instructions of varying specificity and multi-turn conversational exchanges. Moreover, each prompt template was executed five times with identical prompt text to assess reproducibility. For each execution, the traces confirmed correct interpretation of the user request, appropriate decomposition of the task into tool calls, correct routing to the relevant MCP server, valid input arguments passed to each tool, correct CPU core allocation as specified by the user, and agentic responses grounded in the evidence returned by the tools. No execution failures, pipeline errors, or instances of hallucinated output were observed under the tested conditions, and all outputs were consistent with the results returned by the underlying computational tools. In the robustness test (prompt B2), the user deliberately provided incorrect information regarding the radionuclide identity and injected activity. The agent identified the discrepancy between the user-supplied values and the DICOM-extracted metadata, correctly prioritised the metadata, informed the user of the inconsistency, and proceeded with the verified values.

## 4. Discussion

In this study, DosimeTron, a prototype agentic system for personalised internal radiation dosimetry in PET/CT, was developed and evaluated. The system accepts natural-language instructions from the clinical expert. No programming, data preprocessing, or manual MC configuration is required. Upon receiving the user's request, the agent autonomously devises an execution strategy and carries it out through calls to 23 tools exposed via four MCP servers. DosimeTron allows users to supply organ-specific biological half-lives, from which effective decay constants and effective half-lives are derived and incorporated into the cumulated activity calculation. Additionally, it provides per-organ dose conversion factors that encode the patient-specific radiation transport independently of any kinetic model, enabling multiplication of these factors against cumulated activities derived from externally defined pharmacokinetic models without repeating the simulation. Another characteristic of the system is that it does not require a GPU, since the entire pipeline can run on CPU-only hardware, making deployment feasible on standard clinical workstations. Computation time is minimised through automatic distribution of the MC simulation across multiple CPU cores. Taken together, the ease of use, dosimetric accuracy, and computational efficiency of DosimeTron suggest that it could facilitate a shift towards personalised radiation dosimetry by enabling patient-specific Monte Carlo methods to enter routine clinical practice.

The majority of PET/CT dosimetric studies [25-29] rely on software such as OLINDA/EXM [4], IDAC-Dose [30], and MIRDcalc [31], which compute organ doses using pre-calculated S-values derived from standardised phantoms. Despite being a straightforward dosimetric method, this population-based approach has a margin of up to 40% error [6], as it cannot account for inter-patient variability in anatomy and radiotracer distribution. Patient-specific MC simulation addresses this limitation by using the patient’s CT as the anatomical model and PET as the activity distribution, enabling accurate radiation transport modelling. General-purpose MC frameworks [16, 32] can achieve high dosimetric accuracy, but their clinical use is limited by complex, multi-step workflows. Although GPU-accelerated approaches can reduce computation times, such hardware is not routinely available in clinical settings. OpenDose3D [22] was recently released as a 3D Slicer [33] plugin offering an interface for internal dosimetry in PET/CT. However, when Monte Carlo simulation is selected as the dosimetry mode, the user must manually enter the administered activity and radionuclide type, the simulation itself must be executed outside 3D Slicer and the results subsequently imported back into the application for dose calculation. Moreover, the reported MC computation time with OpenDose3D is approximately seven hours per patient study, rendering its application unrealistic in real-life clinical environments. In contrast, DosimeTron fully automates the dosimetry pipeline, dynamically parallelises MC simulations across available CPU cores, and achieves approximately a 15-fold reduction in simulation time compared to OpenDose3D while maintaining equivalent dosimetric accuracy.

The availability of accurate and fast patient-specific MC dosimetry has direct clinical relevance for patient populations in whom cumulative radiation exposure is a concern. Patients with oncological conditions or chronic diseases requiring repeated PET/CT examinations over the course of their diagnostic follow-up, may accumulate non-negligible organ/tissue radiation doses. In such cases, a system like DosimeTron can be deployed to quantify the radiation exposure to individual organs/tissues at each examination, support clinical decision-making regarding scan frequency and administered activity, and enable individualised radiation risk assessment across the patient's imaging history. Beyond individual patient management, DosimeTron can potentially be used for large-scale generation of accurate, patient-specific three-dimensional dose distributions, a resource that is currently unavailable in the literature. When linked with individual patient records, clinical characteristics, comorbidities, and long-term follow-up data, these distributions could serve as the dosimetric substrate for epidemiological studies aimed at quantifying the relationship between organ-absorbed radiation dose from diagnostic radiopharmaceuticals and the incidence of radiation effects, including radiation-induced carcinogenesis and cardiovascular toxicity [34, 35].

This study has several limitations. First, the agentic system was evaluated using only two radionuclides ($^{18}F$ and $^{68}Ga$). While the used dataset served as a representative case example, the system is designed to be extensible and can be easily adapted to accommodate a broader range of used radionuclides for both diagnostic and therapeutic PET/CT applications. Finally, the core reasoning engine of DosimeTron is GPT-5.2, accessed via API calls. As smaller language models continue to improve in reasoning and tool-calling capabilities, fully local deployment of such agentic systems will become feasible.

## 5. Conclusion

This study introduces agentic AI to the field of personalised radiation dosimetry in PET/CT. DosimeTron is a prototype agentic system that demonstrates the capacity of LLM-based agents to autonomously execute complex, multi-disciplinary workflows at the intersection of medical image processing, nuclear medicine, and radiation physics. By enabling clinical experts to interact with a state-of-the-art MC dosimetry pipeline through natural language, without requiring programming expertise or technical knowledge of simulation software, DosimeTron lowers the barrier to patient-specific dosimetry and opens a realistic pathway for its integration into routine clinical practice.

**Acknowledgements**

None.

**Funding**

None.

**Table 1. Organ/tissue dosimetric quantities for $^{18}F$ cases. Values are reported as mean ± SD across studies.**

| Organ / Tissue | Ḋ (µGy/s) | D_scan (mGy) | D_inj (mGy) | DCF (Gy/(Bq·s)) | σ_MC (%) |
|---|---|---|---|---|---|
| Right Kidney | 1.54 | 14.7 ± 6.1 | 22.5 ± 9.1 | $1.13\times10^{-14}$ | 0.04 |
| Left Kidney | 1.38 | 13.1 ± 6.0 | 20.0 ± 9.0 | $1.01\times10^{-14}$ | 0.05 |
| Gallbladder | 1.69 | 16.1 ± 13.4 | 24.8 ± 20.7 | $1.26\times10^{-14}$ | 0.18 |
| Liver | 1.31 | 12.4 ± 3.8 | 19.0 ± 5.8 | $9.62\times10^{-15}$ | 0.01 |
| Spleen | 1.06 | 10.1 ± 3.8 | 15.4 ± 5.6 | $7.78\times10^{-15}$ | 0.05 |
| Duodenum | 0.98 | 9.3 ± 3.3 | 14.2 ± 4.9 | $7.21\times10^{-15}$ | 0.10 |
| Right Lung (Middle Lobe) | 1.05 | 10.0 ± 5.0 | 15.2 ± 7.3 | $7.74\times10^{-15}$ | 0.09 |
| Urinary Bladder | 0.42 | 4.0 ± 3.4 | 6.2 ± 5.1 | $3.27\times10^{-15}$ | 0.11 |
| Right Lung (Lower Lobe) | 0.88 | 8.4 ± 5.2 | 12.8 ± 7.5 | $6.46\times10^{-15}$ | 0.05 |
| Right Adrenal Gland | 0.79 | 7.5 ± 2.2 | 11.5 ± 3.1 | $5.78\times10^{-15}$ | 0.36 |
| Right Lung (Upper Lobe) | 0.71 | 6.8 ± 3.1 | 10.3 ± 4.5 | $5.23\times10^{-15}$ | 0.07 |
| Pancreas | 0.69 | 6.5 ± 2.2 | 10.0 ± 3.2 | $5.02\times10^{-15}$ | 0.10 |
| Left Lung (Lower Lobe) | 0.59 | 5.6 ± 2.7 | 8.6 ± 4.0 | $4.33\times10^{-15}$ | 0.06 |
| Left Lung (Upper Lobe) | 0.61 | 5.8 ± 2.8 | 8.8 ± 4.1 | $4.47\times10^{-15}$ | 0.07 |
| Left Adrenal Gland | 0.57 | 5.4 ± 1.7 | 8.2 ± 2.4 | $4.14\times10^{-15}$ | 0.38 |
| Small Bowel | 0.52 | 4.9 ± 1.4 | 7.6 ± 2.1 | $3.85\times10^{-15}$ | 0.04 |
| Stomach | 0.55 | 5.2 ± 1.9 | 8.0 ± 2.8 | $4.04\times10^{-15}$ | 0.09 |
| Esophagus | 0.44 | 4.1 ± 1.0 | 6.3 ± 1.4 | $3.19\times10^{-15}$ | 0.17 |
| Colon | 0.41 | 3.9 ± 1.3 | 6.0 ± 2.0 | $3.06\times10^{-15}$ | 0.06 |
| Heart | 0.42 | 4.0 ± 1.1 | 6.1 ± 1.3 | $3.08\times10^{-15}$ | 0.03 |
| Spinal Cord | 0.31 | 3.0 ± 1.0 | 4.5 ± 1.4 | $2.26\times10^{-15}$ | 0.11 |
| Thyroid | 0.26 | 2.5 ± 0.8 | 3.8 ± 1.2 | $1.93\times10^{-15}$ | 0.32 |

Ḋ: dose rate at scan acquisition time (µGy/s); D_scan: absorbed dose, scan-to-infinity integration (mGy); D_inj: absorbed dose, injection-to-infinity integration (mGy); DCF: dose conversion factor (Gy/(Bq·s)); σ_MC: mean Monte Carlo statistical uncertainty (%).

**Table 2. Organ/tissue dosimetric quantities for $^{68}$Ga cases. Values are reported as mean ± SD across studies.**

| Organ / Tissue | $\dot{D}$ (μGy/s) | D_scan (mGy) | D_inj (mGy) | DCF (Gy/(Bq·s)) | σ_MC (%) |
|---|---|---|---|---|---|
| Right Kidney | 4.07 | 23.8 ± 10.4 | 46.6 ± 19.5 | $5.19\times10^{-14}$ | 0.03 |
| Left Kidney | 3.84 | 22.5 ± 10.1 | 44.2 ± 19.2 | $4.94\times10^{-14}$ | 0.03 |
| Gallbladder | 0.45 | 2.7 ± 1.3 | 5.2 ± 2.4 | $5.70\times10^{-15}$ | 0.27 |
| Liver | 0.85 | 5.0 ± 2.1 | 9.7 ± 3.5 | $1.07\times10^{-14}$ | 0.02 |
| Spleen | 1.16 | 6.8 ± 3.3 | 13.1 ± 5.7 | $1.45\times10^{-14}$ | 0.04 |
| Duodenum | 1.15 | 6.7 ± 2.8 | 13.2 ± 5.3 | $1.46\times10^{-14}$ | 0.08 |
| Right Lung (Middle Lobe) | 0.70 | 4.1 ± 2.4 | 7.9 ± 4.3 | $8.93\times10^{-15}$ | 0.07 |
| Urinary Bladder | 1.58 | 9.3 ± 7.3 | 18.8 ± 15.7 | $2.12\times10^{-14}$ | 0.05 |
| Right Lung (Lower Lobe) | 0.66 | 3.9 ± 2.1 | 7.6 ± 3.8 | $8.62\times10^{-15}$ | 0.04 |
| Right Adrenal Gland | 0.63 | 3.7 ± 1.7 | 7.1 ± 3.0 | $7.90\times10^{-15}$ | 0.36 |
| Right Lung (Upper Lobe) | 0.57 | 3.3 ± 1.6 | 6.5 ± 2.7 | $7.26\times10^{-15}$ | 0.05 |
| Pancreas | 0.54 | 3.2 ± 1.2 | 6.1 ± 2.0 | $6.85\times10^{-15}$ | 0.09 |
| Left Lung (Lower Lobe) | 0.56 | 3.3 ± 1.5 | 6.4 ± 2.9 | $7.32\times10^{-15}$ | 0.05 |
| Left Lung (Upper Lobe) | 0.50 | 2.9 ± 1.3 | 5.6 ± 2.2 | $6.30\times10^{-15}$ | 0.05 |
| Left Adrenal Gland | 0.51 | 3.0 ± 1.3 | 5.8 ± 2.3 | $6.52\times10^{-15}$ | 0.34 |
| Small Bowel | 0.54 | 3.1 ± 1.3 | 6.1 ± 2.3 | $6.76\times10^{-15}$ | 0.03 |
| Stomach | 0.41 | 2.4 ± 1.1 | 4.7 ± 2.0 | $5.22\times10^{-15}$ | 0.07 |
| Esophagus | 0.40 | 2.4 ± 0.9 | 4.6 ± 1.4 | $5.06\times10^{-15}$ | 0.16 |
| Colon | 0.35 | 2.0 ± 0.8 | 3.9 ± 1.4 | $4.37\times10^{-15}$ | 0.04 |
| Heart | 0.31 | 1.8 ± 0.6 | 3.5 ± 1.0 | $3.82\times10^{-15}$ | 0.04 |
| Spinal Cord | 0.25 | 1.5 ± 0.9 | 2.9 ± 1.6 | $3.10\times10^{-15}$ | 0.11 |
| Thyroid | 0.25 | 1.4 ± 0.6 | 2.8 ± 1.1 | $3.07\times10^{-15}$ | 0.32 |

$\dot{D}$: dose rate at scan acquisition time (μGy/s); D_scan: absorbed dose, scan-to-infinity integration (mGy); D_inj: absorbed dose, injection-to-infinity integration (mGy); DCF: dose conversion factor (Gy/(Bq·s)); σ_MC: mean Monte Carlo statistical uncertainty (%).

**Table 3. Scan-time dose rate comparison between the DosimeTron pipeline and OpenDose3D. Values are mean ± SD (µGy/s).**

| Organ / Tissue | OpenDose3D (µGy/s) | DosimeTron (µGy/s) | Bias (µGy/s) | 95% LoA (µGy/s) | r | CCC | MAPD (%) |
|---|---|---|---|---|---|---|---|
| Right Kidney | 2.35 ± 1.59 | 2.33 ± 1.57 | -0.019 | [-0.088, +0.050] | 1.000 | 1.000 | 1.0 |
| Left Kidney | 2.16 ± 1.60 | 2.15 ± 1.58 | -0.013 | [-0.145, +0.119] | 0.999 | 0.999 | 1.4 |
| Gallbladder | 1.40 ± 1.34 | 1.37 ± 1.28 | -0.022 | [-0.234, +0.189] | 0.997 | 0.996 | 4.2 |
| Liver | 1.22 ± 0.47 | 1.22 ± 0.47 | -0.001 | [-0.014, +0.012] | 1.000 | 1.000 | 0.4 |
| Spleen | 1.10 ± 0.47 | 1.10 ± 0.47 | -0.003 | [-0.072, +0.065] | 0.997 | 0.997 | 1.2 |
| Duodenum | 1.04 ± 0.42 | 1.03 ± 0.42 | -0.011 | [-0.081, +0.060] | 0.997 | 0.996 | 2.6 |
| Right Lung (Middle Lobe) | 0.94 ± 0.49 | 0.97 ± 0.51 | +0.030 | [-0.044, +0.105] | 0.997 | 0.995 | 4.1 |
| Right Lung (Lower Lobe) | 0.81 ± 0.41 | 0.83 ± 0.42 | +0.021 | [-0.028, +0.070] | 0.999 | 0.997 | 2.7 |
| Right Adrenal Gland | 0.77 ± 0.28 | 0.78 ± 0.28 | +0.005 | [-0.115, +0.125] | 0.976 | 0.976 | 5.9 |
| Urinary Bladder | 0.76 ± 1.10 | 0.75 ± 1.07 | -0.011 | [-0.166, +0.144] | 0.998 | 0.997 | 2.4 |
| Right Lung (Upper Lobe) | 0.67 ± 0.29 | 0.70 ± 0.31 | +0.029 | [-0.024, +0.083] | 0.998 | 0.991 | 4.2 |
| Pancreas | 0.65 ± 0.19 | 0.65 ± 0.20 | +0.001 | [-0.037, +0.038] | 0.995 | 0.995 | 2.0 |
| Left Lung (Upper Lobe) | 0.59 ± 0.26 | 0.61 ± 0.27 | +0.021 | [-0.015, +0.058] | 0.998 | 0.994 | 3.7 |
| Left Lung (Lower Lobe) | 0.59 ± 0.26 | 0.60 ± 0.27 | +0.014 | [-0.026, +0.054] | 0.998 | 0.996 | 2.5 |
| Left Adrenal Gland | 0.57 ± 0.19 | 0.57 ± 0.20 | +0.006 | [-0.097, +0.108] | 0.965 | 0.963 | 5.2 |
| Small Bowel | 0.53 ± 0.15 | 0.52 ± 0.15 | -0.004 | [-0.044, +0.037] | 0.990 | 0.990 | 1.8 |
| Stomach | 0.51 ± 0.19 | 0.52 ± 0.19 | +0.004 | [-0.039, +0.046] | 0.994 | 0.994 | 2.4 |
| Esophagus | 0.43 ± 0.11 | 0.43 ± 0.11 | +0.001 | [-0.050, +0.052] | 0.972 | 0.972 | 3.5 |
| Colon | 0.41 ± 0.14 | 0.41 ± 0.14 | -0.002 | [-0.033, +0.029] | 0.994 | 0.994 | 2.1 |
| Heart | 0.40 ± 0.12 | 0.40 ± 0.12 | +0.001 | [-0.005, +0.008] | 1.000 | 1.000 | 0.7 |
| Spinal Cord | 0.30 ± 0.13 | 0.31 ± 0.13 | +0.003 | [-0.014, +0.020] | 0.998 | 0.997 | 2.1 |
| Thyroid | 0.26 ± 0.10 | 0.26 ± 0.10 | -0.002 | [-0.035, +0.030] | 0.986 | 0.986 | 5.5 |

Bias: Bland-Altman mean difference (DosimeTron - OpenDose3D); 95% LoA: limits of agreement; r: Pearson correlation coefficient; CCC: Lin's concordance correlation coefficient; MAPD: mean absolute percentage difference.

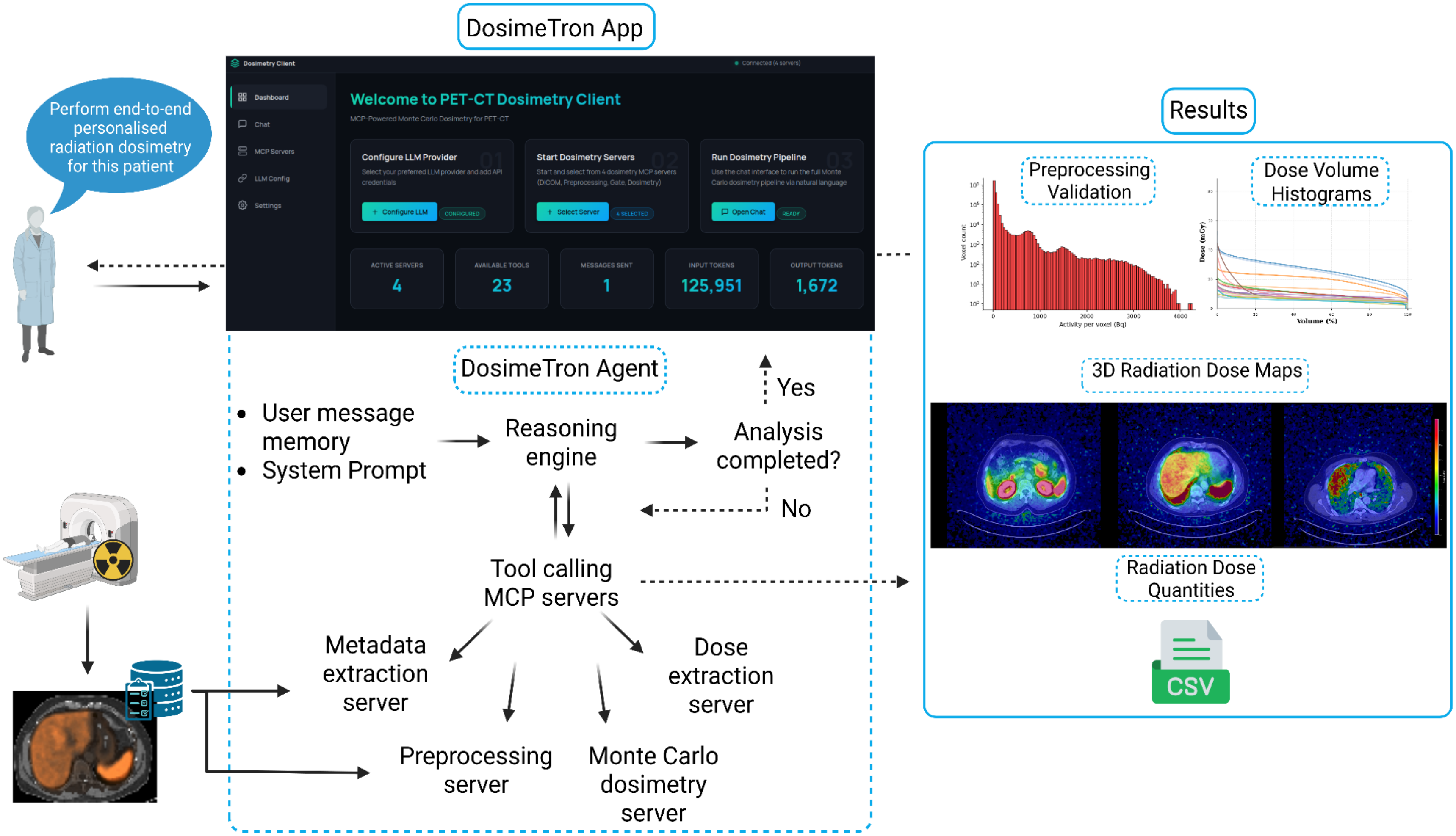

**Figure 1**. Flowchart of the pipeline for deriving personalised internal dosimetry with DosimeTron. The user requests end-to-end patient-specific radiation dosimetry, and the DosimeTron agent analyses the task, considers the computational tools exposed by the MCP servers, devises a solution strategy, and executes it by selecting and using the appropriate tools, resulting in quantitative and qualitative dosimetric outputs. MCP: Model Context Protocol.

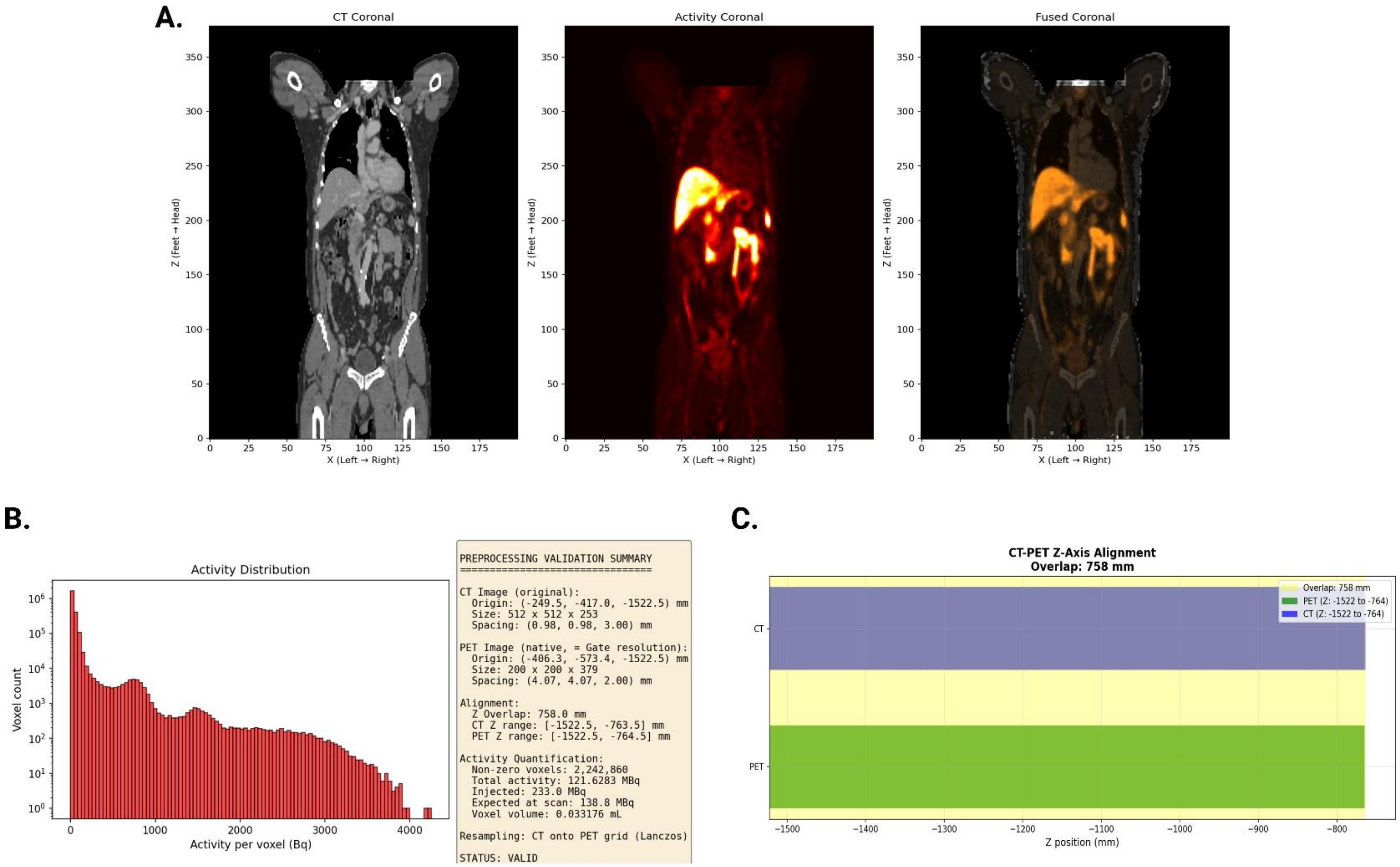

**Figure 2**. Preprocessing verification data. In the initial preprocessing steps the system aligns and registers CT to PET images and generates a series of visualization data for the user to inspect and confirm appropriate alignment. This figure demonstrates the correct alignment of CT, PET and fused images (A), the activity distribution histogram and preprocessing validation report (B) as well as the spatial alignment of CT and PET images along the Z-axis (C) for an indicative patient.

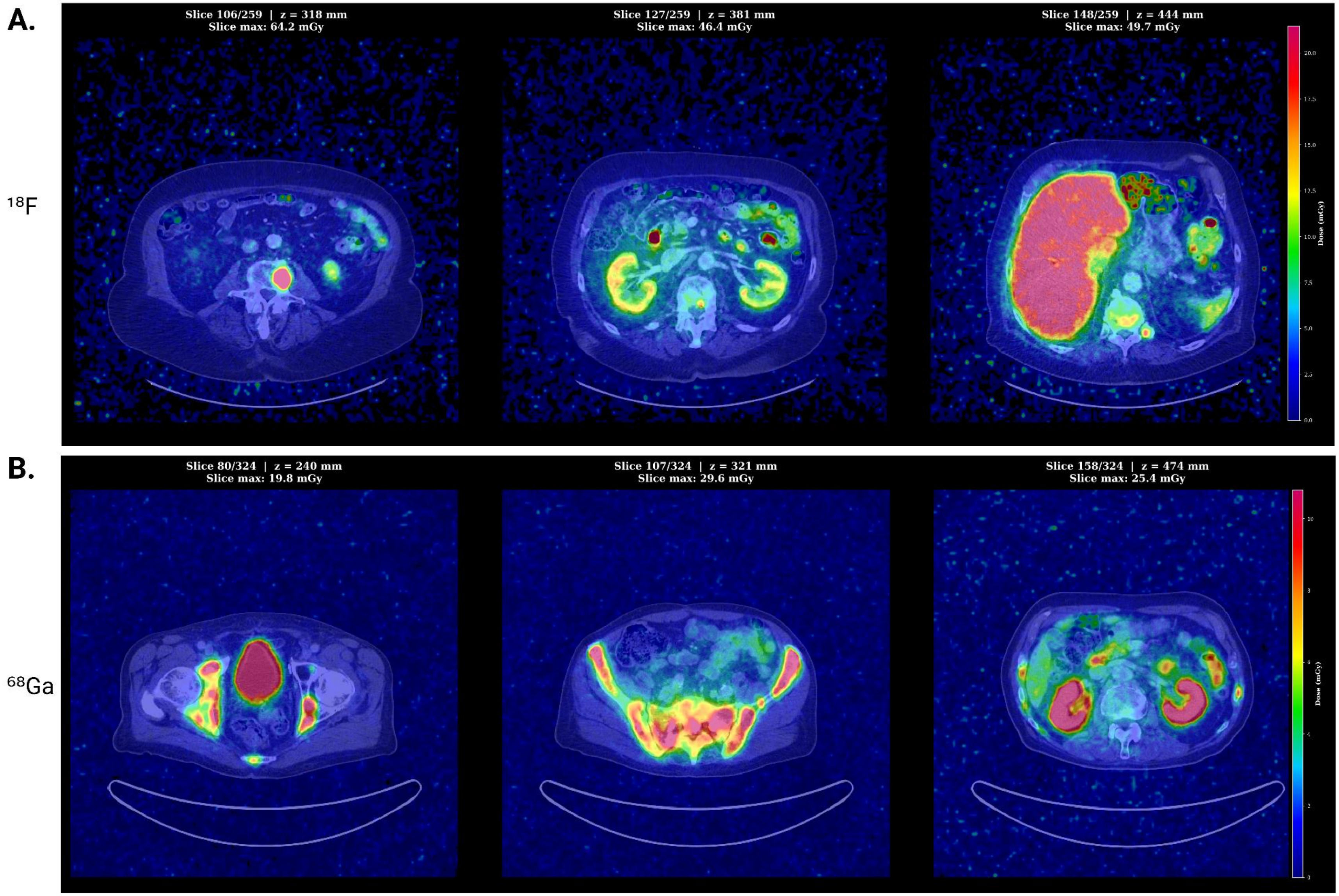

**Figure 3**. Indicative radiation dose slices of 3D dose maps generated for two PET/CT scans, one with 18F-PSMA (A) and a second with 68Ga-PSMA (B).

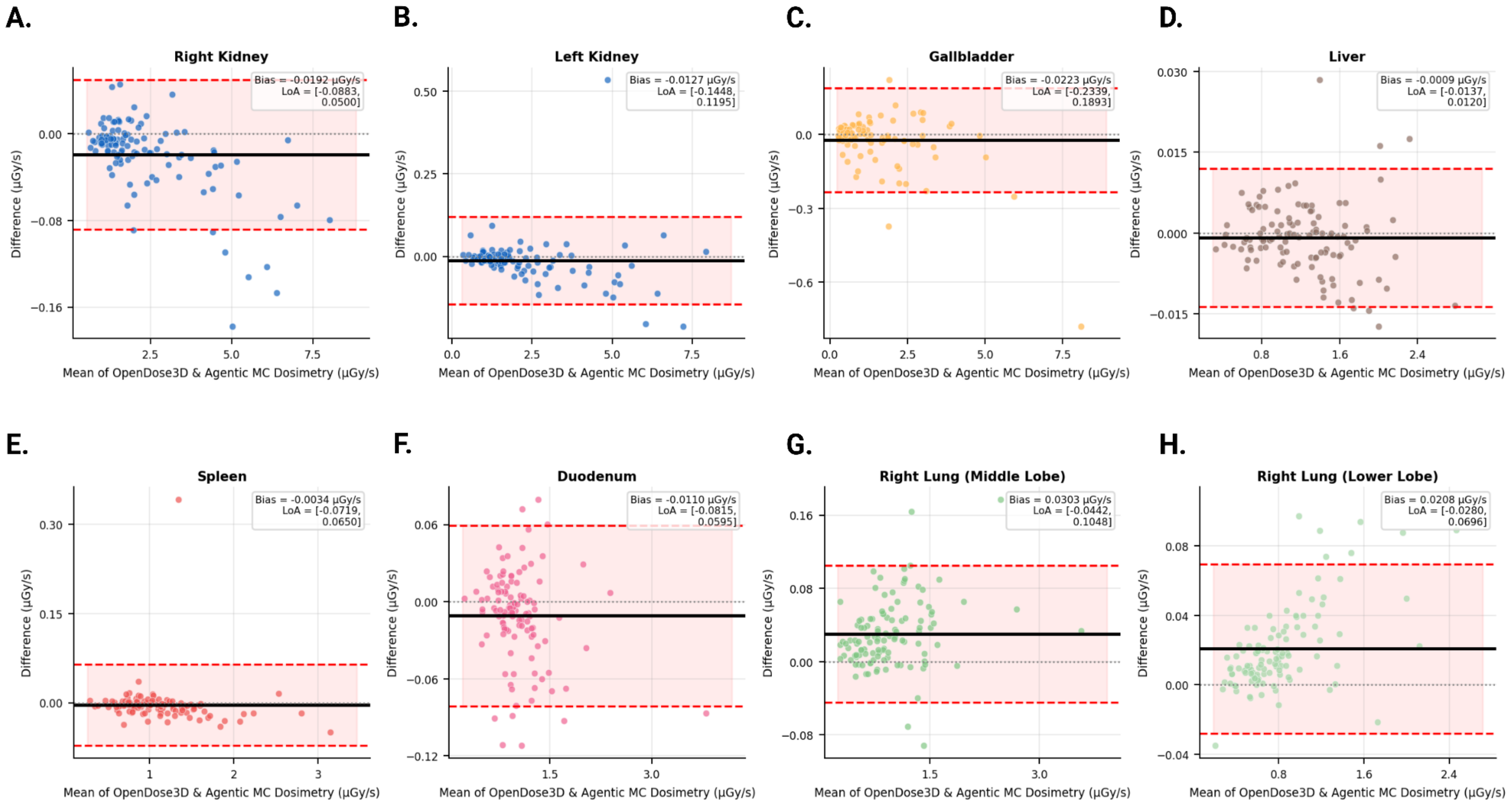

**Figure 4**. Agreement between DosimeTron and OpenDose3D. Bland-Altman plots for the 8 anatomical structures receiving the highest radiation dose. MC: Monte Carlo; LoA: limit of agreement.

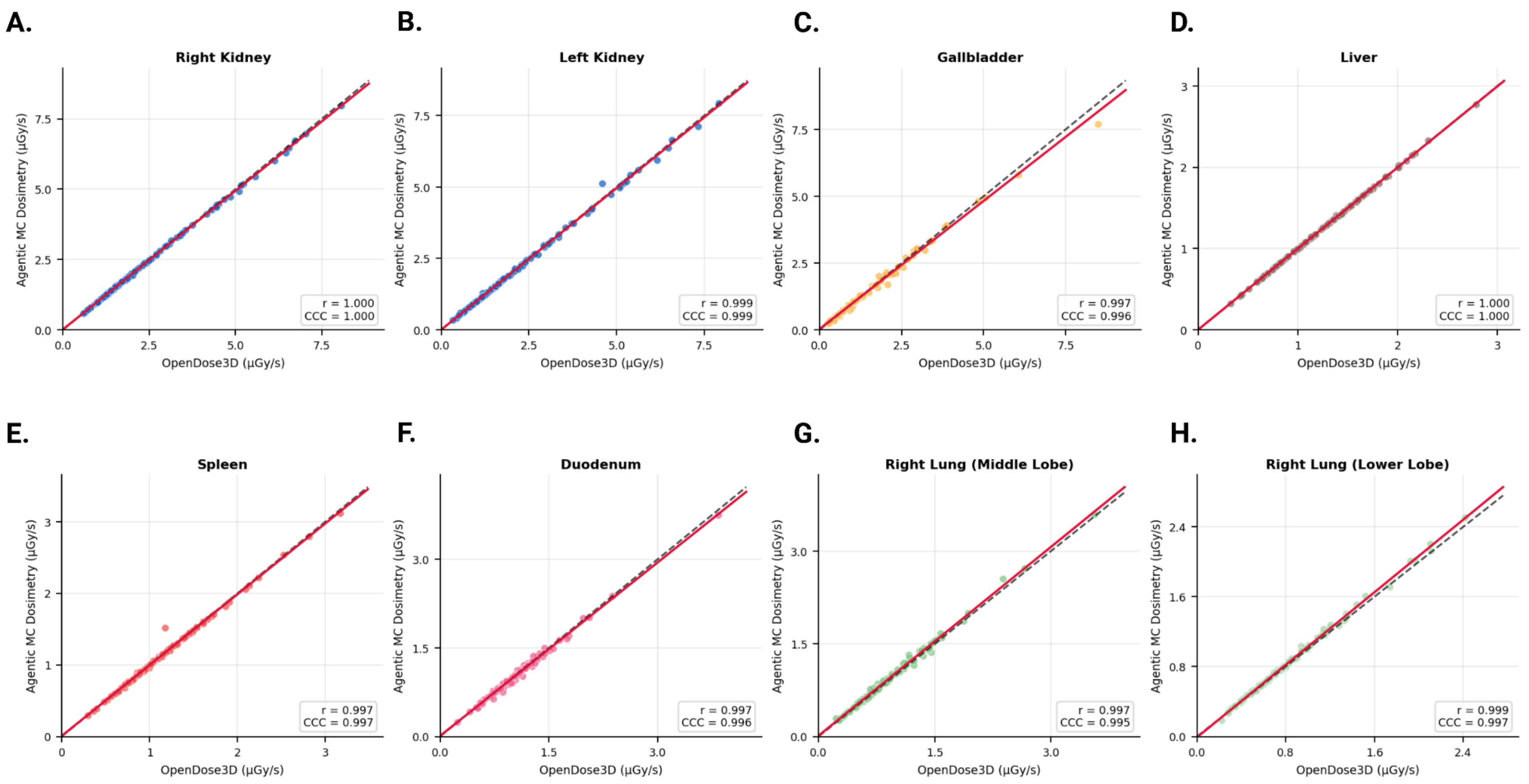

**Figure 5**. Scatter plots comparing dose rates (µGy/s) calculated with DosimeTron and OpenDose3D for the 8 anatomical structures receiving the highest radiation dose. MC: Monte Carlo; CCC: concordance correlation coefficient; r: Pearson's correlation coefficient